\documentclass[12pt,preprint]{aastex}

\usepackage{CJK}

\usepackage{epsfig}

\shorttitle{A Halo of NGC\,7293}
\shortauthors{Zhang, Hsia \& Kwok}

\begin{document}


\begin{CJK*}{Bg5}{bsmi}
\CJKtilde

\title{DISCOVERY OF A HALO AROUND THE HELIX NEBULA NGC\,7293 IN THE {\it WISE} ALL-SKY SURVEY}

\author{Yong Zhang (±iªa), Chih-Hao Hsia (®L§Ó¯E), Sun Kwok (³¢·s)}

\affil{Department of Physics, Faculty of Science, The University of Hong Kong, Hong Kong, China
\email{zhangy96@hku.hk,xiazh@hku.hk,sunkwok@hku.hk}}

\begin{abstract}
We report the discovery of an extended halo ($\sim40'$ in diameter) around the planetary nebula NGC\,7293 (the Helix Nebula) observed in 12\,$\mu$m band from the {\it Wide-field Infrared Survey Explorer}  all-sky survey. The mid-infrared halo has an axisymmetric structure with a sharp 
boundary to the northeast and a more diffuse boundary to the southwest, suggesting 
an interaction between the stellar wind and the interstellar medium (ISM). The symmetry axis of the halo is well aligned with that of a northeast arc, suggesting that the two structures are physically associated. 
We have attempted to fit the observed geometry with a model of a moving steady-state stellar wind interacting with the ISM.  Possible combinations of the ISM density and the stellar velocity are derived from these fittings.   The discrepancies between the model and the observations suggest that the stellar mass loss has a more complicated history, including possible time and angle dependences.

\end{abstract}


\section{INTRODUCTION}

As low- and intermediate-mass stars evolve from asymptotic giant branch (AGB)
to planetary nebula (PN) stages, they continually eject matter into the
interstellar medium (ISM). The resultant circumstellar envelopes
are one of the main sources of chemical enrichment of the Galaxy.
Our current understanding of the origin of PNs is based on the interacting stellar wind (ISW) model \citep{kwo82} which has predicted the presence of AGB halos created by AGB winds.  The remnants of this AGB wind are swept up by a later-developed fast wind to form a PN.   One of the predictions of the model is that the remnants of the AGB wind should be detectable as halos in PNs.  Since then, AGB halos have been detected by a variety of techniques.  These include observations (1) in the visible through emission or scattering, (2) in millimeter and submillimeter wavelengths through molecular-line emission, and (3) in the infrared through dust emission.  Since PN halos can contain more mass than the main bright nebular shells, studies of PN halos are important to understand the AGB mass-loss history and the chemical evolution of the Galaxy.

In the visible, AGB halos are usually very faint (with surface brightness about $10^{-3}$--$10^{-4}$ lower than the main nebulae), making them difficult to observe.  Many efforts have been made to search for faint AGB halos \citep[see e.g.][]{cor03,ram09}. These observations show that the detected AGB halos usually exhibit a bow-shaped bright rim, suggesting the interaction between AGB winds and the ISM. The wind-ISM interaction process has been the subject of various theoretical studies 
\citep[see e.g.][]{smi76,bor90,sok91,dga98,vil03,war07,war07b}. Early studies suggest that the wind-ISM interaction is observable only for evolved PNs whose density has dropped below a critical limit.  More recent hydrodynamic simulations argue that the interaction can be observed for young PNs if the time-dependent variations in AGB and post-AGB winds are taken into account.   
The wind-ISM interaction was indeed observed in not only ancient PNs \citep{twe96} but also in the circumstellar envelopes of  AGB stars \citep{uet06,uet08,cox12}.

The high-quality images acquired by recent wide field-of-view (FOV) surveys have been proven useful to search for large-scale extended structures around evolved stars. For example, \citet{sah10} discovered a large outer shell around the highly evolved AGB star IRC+10216 using the images taken with {\it Galaxy Evolution Explorer} \citep[{\it GALEX};][]{mar05}.
In this paper, we report the results of searching for outer structures around NGC\,7293 utilizing the recently released mid-infrared images of the {\it Wide-field Infrared Survey Explorer} \citep[{\it WISE};][]{wri10} all-sky survey.

NGC 7293 is one of the most extensively studied PNs \citep[see][and reference therein]{ode04,hor06,spe02,mea92,mea05,mea08}. Given its relative proximity to the Sun and its relatively accurate distance measurement, NGC\,7293 is an excellent target to study in detail the spatial structure of PNs.  In this paper,  we have adopted the distance estimate of $216^{+14}_{-12}$\,pc as determined by \citet{ben09}.
Although its overall appearance seems simple, a number of intriguing structures, from small cometary knots to large-scale  arcs, to bipolar outflows, have been revealed in previous observations. The inner helical structure is composed  of thousands of cometary knots of  lowly-ionized and molecular gas. The main nebula consists of two rings of highly-ionized gas and a faint outer filament.
The three-dimensional (3-D) structure of the main nebula has been investigated by \citet{mea08} and \citet{ode04} who suggested that it was formed through multiple outflows along different directions.

Although NGC\,7293 is of great astrophysical interest, it is not an easy object to observe because of its low surface brightness and its large angular extent in the sky.  Without highly sensitive and large FOV observations, it is impossible to explore the faint halo around this PN.  The recent {\it WISE} observations, therefore, provide an invaluable resource for that purpose.

\section{The Data}

The  data used in this paper were taken from the {\it WISE} All-Sky Data Release\footnote{see 
http://wise2.ipac.caltech.edu/docs/release/allsky/}.
The data have been calibrated with the {\it WISE}-pipeline.
{\it WISE} mapped the  entire sky in four bands centered at wavelengths of 3.4, 4.6, 12, and 22\,$\mu$m
(W1--W4) with an angular resolution of 6.1$''$, 6.4$''$, 6.5$''$, and 12.0$''$,
respectively. During the survey,  the four bands were simultaneously deployed to image a $47'\times47'$ FOV on the sky.  The exposure time of individual frame is 7.7 seconds in W1 and W2, and 8.8 seconds in W3 and W4.  The number of exposures are different at different ecliptic latitudes.
The observations of the fields including NGC\,7293 were made on
 20--23 May 2010, with total accumulated time of 17.2 minutes for W1 and W2, and 19.7 minutes for W3 and W4.  The composite-color {\it WISE} image of NGC 7293 is shown in Fig.~\ref{f1}.

Our analysis is supplemented by the far- and near-ultraviolet (FUV and NUV) images retrieved from the {\it GALEX} archive. The FUV and NUV images have effective wavelengths of 1516\,{\AA} and 2267\,{\AA}, and
angular resolutions of 4.5$''$ and 6$''$, respectively.
A detailed description of the calibration and
data products of {\it GALEX} can be found in \citet{mor07}.
The observations of NGC\,7293 were made on 22--25 August 2004 as part of the 
Calibration Imaging Survey (CAI). The accumulated exposure time is 56 minutes for each band.
In order to improve the signal-to-noise ratio, we co-added the FUV and NUV images.
The resultant {\it GALEX} image is shown in the right panel of Fig.~\ref{f1}.

\section{RESULTS}

\begin{figure*}
\epsfig{file=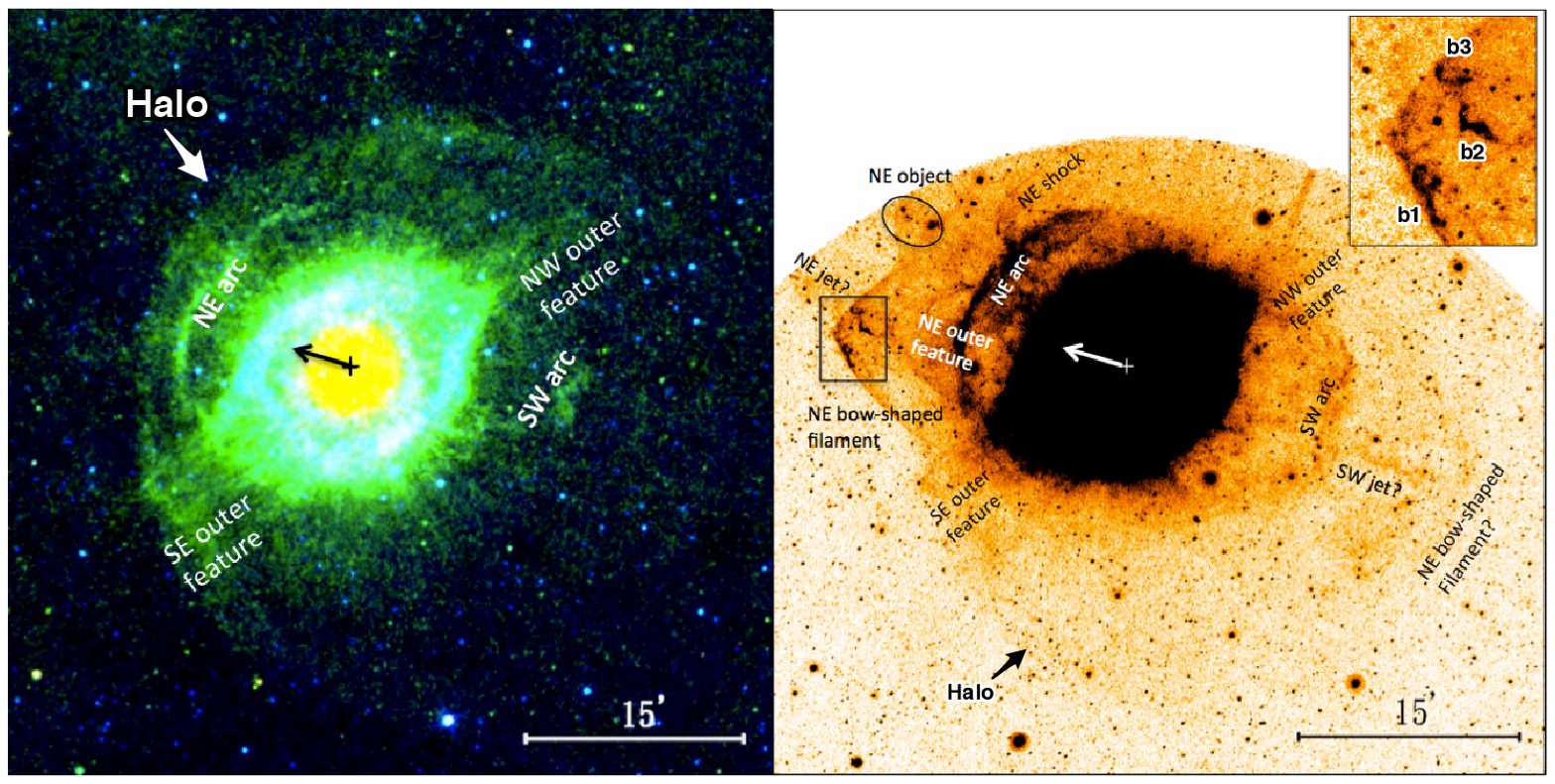, height=8.2cm}
\caption{Composite-color {\it WISE} image (left panel)
and false-color {\it GALEX} image (right panel) of NGC\,7293. 
Left is east, and up is north.
In the left panel, the WISE 3.4\,$\mu$m band is assigned the color blue, the 12\,$\mu$m band the green, and the 22\,$\mu$m band the red.  The labels ``NW outer feature'' and ``SE outer feature'' are the terms assigned by \citet{ode04} for the bipolar outflow.
In the right panel, the top-right corner panel shows a zoomed-in view of the black box.
The cross and arrow at the center of each panel represent the location of
the central star and the direction of proper motion. An axisymmetric
halo is clearly visible in the {\it WISE} image. Note that
in the {\it GALEX} image, the spike above the NW outer feature is artificial.
}
\label{f1}
\end{figure*}

\begin{figure*}
\centering
\epsfig{file=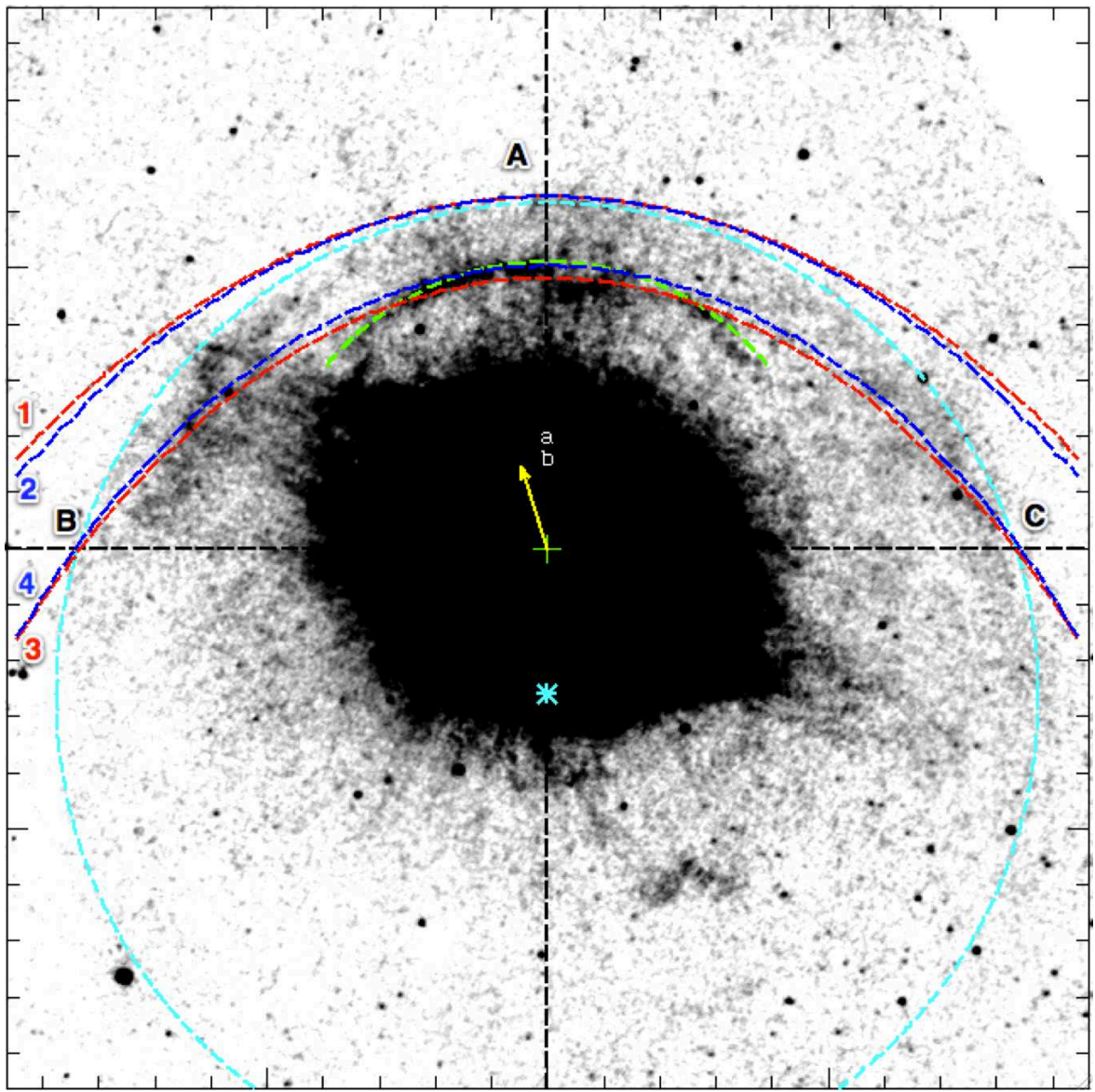, height=15.5cm}
\caption{{\it WISE} 12\,$\mu$m image shown in gray scale. The area of the image is
$44'\times44'$ with the north direction $60^\circ$ clockwise from the vertical axis. 
The cross and arrow represent the location of the central star and the direction of proper motion (as in Fig.~\ref{f1}).
The boundary of the halo is indicated by the cyan dashed circle with a radius of $20'$ from its center (denoted by a cyan colored asterisk).  The position of the NE arc is indicated by the green dashed circular arc
with a radius of $11.7'$ from the central star. 
The red and blue dashed curves represent the four models listed in Table~\ref{t1} and the number of the model is labeled next to the curve.   The projected locations of the apexes of models 2 and 4 are marked by  labels ``a'' and ``b''.}
\label{model}
\end{figure*}

An axisymmetric extended halo at 12\,$\mu$m surrounding NGC 7293 can clearly be seen in Fig.~\ref{f1}. 
The halo exhibits a bow-like shape with a well-defined boundary to the northeast (NE), in contrast to a diffuse and open structures  to the southwest (SW).  Such asymmetric morphology gives a clear indication that the object is moving toward the NE.  The direction of motion can be approximated by the symmetry axis of the halo and the symmetry axis is found to located at a  position angle (PA) of $60^\circ$. The NE boundary  of the halo can be approximately described by a circle
with a radius of $1213''$ ($3.9\times10^{13}$\,km$=2.6\times10^{5}$\,AU$=1.3$\,pc), with a center located at $359''$ ($1.2\times10^{13}$\,km$=7.8\times10^{4}$\,AU$=0.4$\,pc) from the central star (see Fig.~\ref{model}). 

We find that  the 12\,$\mu$m emission in the halo is relatively smooth and is about 10 times fainter than that of the main nebula.
This halo is not detectable at the {\it WISE} 22\,$\mu$m band and has not been seen in previously reported  optical images\footnote{After this paper was published, we were informed by Q. Parker that their deep H$\alpha$ image has revealed
some faint extended structures that are associated with the halo (Parker et al. 2001).}. 
 It is only marginally detectable in the {\it WISE} 3.4 and 4.6\,$\mu$m and in the {\it GALEX} images. 
 Assuming that the infrared emission peaks at 12 $\mu$m, we infer that the emission of the halo is dominated by thermal emission from warm dust with a temperature of $\sim300$\,K. One can in principle obtain the spatial distribution of temperature in the halo by comparing the W1--W3 images. However, the halos in W1 and W2 are too faint to yield meaningful conclusions. Nevertheless, we cannot rule out the possibility that shock-excited H$_2$ lines might significantly contribute to the mid-infrared emission of the halo, as suggested by \citet{ram10} in their study of the halo of NGC\,40.

The shape of the halo obviously suggests that it is interacting with the ISM.
The conclusion is strengthened by the study of the proper motion of the Helix Nebula by \citet{ker08}.   The direction of the proper motion is found to be at PA$=77^\circ$, which roughly aligns with the symmetry axis of the halo and the bow apex. The apex of the bow-shaped halo (position $A$ in Fig.~\ref{model}) is about $854''$ ($2.8\times10^{13}$\,km=$1.8\times10^{5}$ AU=0.9 pc)
from the central star.  Assuming the velocity of AGB wind, $v_{\rm w}=10$\,km\,s$^{-1}$,
we obtain a lower limit to the dynamic age of the halo of 88\,000\,yr, suggesting that
the halo was indeed formed long ago, and can therefore serve as a fossil record of
the AGB mass-loss history. \citet{ode04} estimated that the outer material of NGC\,7293 has a recombination timescale of $\sim40\,000$\,yr, much shorter than the dynamic age of the halo. Therefore, it is very likely that the halo is neutral and optically invisible.
Based on optical observations,  \citet{mea08} determined the
dynamic age of the main nebula to be 11\,000\,yr.  If the halo has been continually developed 
with a mass-loss rate of $\dot M=10^{-5}$\,$M_\sun$\,yr$^{-1}$ and was terminated when a fast wind was launched to form the main nebula, the mass of the halo can be estimated to be $>$0.77\,$M_\sun$.  Due to the wind-ISM interaction, the past of the halo material has probably been stripped to the SW behind the central star.

The {\it WISE} images also reveal some smaller-scale structures, such as the NE arc and
small cometary knots. The main nebula exhibits a collimated bipolar outflow oriented NW--SE (the NW and SE outer features in Fig. 1, as assigned by O'Dell et al. 2004).  We note that the SE ends of the bipolar outflow  are running into the boundary of the halo. These results are consistent
with the narrower FOV infrared observations acquired with the {\it Spitzer Space Telescope} by 
\citet{hor06}.  The NE arc is the brightest feature outside the main nebula.
Fig.~\ref{model} shows that the 12\,$\mu$m image of the
NE arc can be described by a circle arc about $696''$ from the central star while its FUV emission is closer to the central star (about $680''$ away). \citet{hor06}
found that the H$_2$ mid-infrared emission of the NE arc is located
in a shell outside the H$\alpha$ emission. 
Because the {\it GALEX} FUV emission is presumably from shock-excited H$_2$, the 
stratified optical/FUV and mid-infrared emission structures
likely suggest the radial variations of excitation conditions rather than the H$^+$/H$_2$ ratios.
We also note that  the {\it WISE} 22\,$\mu$m image shows strong brightening
around the central star. This confirms the previous {\it Spitzer} observations at 24\,$\mu$m
by \citet{su07} who ascribed the 24\,$\mu$m brightening to the presence of a inner debris disk.

In the outer region, the {\it GALEX} and {\it WISE} images appear strikingly
different (see Fig.~\ref{f1}). The {\it GALEX} image shows many bow-shaped features that are not detected by {\it WISE}.  A schematic presentation of the observed structures is presented in Fig.~6 of  \citet{mea08}.
These FUV features are presumably due to H$_2$ emission excited by fast shock waves. The most interesting feature is the NE bow-shaped filament, which contains three bow-shaped sub-structures ($b1$--$b3$ in the inset of Fig.~\ref{f1}) and whose apex is exactly aligned with the direction of proper motion (see the proper motion vector in Fig.~\ref{f1}).  These sub-structures were probably formed due to instabilities in the shock interface, as predicted by the simulations of \citet{moh12}. 
We note that the apex of the fast outflow shown in the {\it GALEX} image has a ``$L$'' shape, which is characteristic of an outflow near the plane of the sky.  If the outflow is inclined at a large angle, the apex would take on a more rounded shape.

Unlike the bow-shaped filament, the NE boundary of the halo cannot be seen in the {\it GALEX} image. This is probably the result of the shock responsible for the halo boundary being too slow to provide enough energy to excite H$_2$ emission. The {\it GALEX} image also shows the NE object and shock \citep[see][for a discussion]{ode04}, as well as bipolar streams in NE-SW orientation (marked as ``outer features'' in Fig.~\ref{f1}). 
Although there are several asymmetrical outflows (outside and inside the halo), the halo boundary seems to be continuous and not broken (as seen in the left panel of Fig.~1). This simply suggests that
from the perspective perpendicular to the line of sight, these streams do not cross the projected halo boundary.

The mid-infrared halo has a appearance similar to the shell around IRC+10216 detected by \citet{sah10},
except that the vortices behind the IRC+10216 shell are not clearly detected in NGC\,7293. 
If the vortices had been formed behind the halo of NGC\,7293, they supposedly
would have been diluted to the ISM.

Overall, the newly discovered mid-infrared halo of the well studied PN NGC\,7293 strengthens the conclusions that AGB halos are commonly present and infrared imaging is particularly useful to detect the faint extended structures \citep[e.g.][]{ram09,ram12}. The direct detection of the halo of NGC 7293 provides us with the knowledge of the ambient environment within which the later-developed dynamical processes [e.g., the multipolar fast outflows \citep{mea08}] operates.  By deriving the geometry and density of the halo, we are in a much better position to constrain the physical processes at work.  

\section{DISCUSSION}

The shape of the 12\,$\mu$m halo suggests the presence of a bow shock.  Since the shape of bow shock is dependent on the velocity and density contrasts between the matter of the stellar wind and the ISM, the observed structure can serve as a useful probe of the properties of the AGB wind and the ISM. 
If the post-shock material has a high cooling rate, the physical thickness of the shell at the shock interface would be small. Under the thin-shell approximation and assuming that  an equilibrium between the ram pressure of stellar wind and that of the ISM is reached, \citet{wil96} deduced an analytic
expression for the shape of the bow shock,

\begin{equation}
R(\theta)=R_0\sin^{-1}\theta \sqrt{3(1-\theta\cot \theta)},
\end{equation}
where $\theta$ is the polar angle from the symmetric axis as seen from the central star at the coordinate origin, and $R_0$ is the distance between the apex and the central star (i.e. stand-off distance) given by
\begin{equation}
R_0=\sqrt{\dot Mv_{\rm w}/(4\pi \rho_{\rm ISM} v_*^2)}.
\end{equation}
In eq.~2, $\rho_{\rm ISM}$ is the mass density of the ISM and $v_*$ the velocity of star respect to the ISM.  $R(\theta)$ increases with increasing $\theta$ because the ram pressure of the ISM
is the strongest at the apex and decreases on either side. Assuming that the properties of the
stellar wind are typical (see Table~\ref{t1}), one can obtain the relation between $v_*$ and the number density  ($n_{\rm ISM}$) of the ISM from eq.~(2) by measuring $R_0$. Wilkin's analytic solutions have been widely applied to study the bow shocks around a variety of objects including young stellar objects, red supergiants, AGB stars, and pulsars \citep[e.g.][]{cox12,mac12,jor11}.

We have attempted to determine the $R_0$ value by fitting the projected shape of a 3-D paraboloid (constructed from eq.~1) on the sky plane with the observations.  For given assumed values of $R_0$ and $i$ (inclination angle of the symmetric axis respect to the sky plane;
 no matter whether the apex points toward or away from us), the projected distances between every point on the paraboloid and  the central star can be determined.  The boundary of the projected 3-D paraboloid is found by calculating the furthermost projected distance at each position angle in the sky plane.  Values of $R_0$ and $i$ are then adjusted to obtain the best fit.
Similar procedures have been used by \citet{cox12} to study the bow shocks around AGB stars and red supergiants.  We should note that since the bow shock has a sharp boundary, the inclination angle cannot be too large (i.e., the star is moving not far from the plane of the sky).

We  constructed a series of models with different $i$ and $R_0$ values, but found none of them can completely account for the shape of the halo boundary. Fig.~\ref{model} shows four of the models and the model parameters are listed in Table~\ref{t1}. We can see that as $i$ increases, the projected location of the apex of the paraboloid changes from $A$ toward the central star (positions $a$ and $b$).   If $i$ were 90\degr, the projected position of the apex would be at the central star.   The models suggest that given a certain distance from the projected apex (position $A$) to the central star, the values of $R_0$ monotonically decreases with increasing $i$ (see models 1 and 2). This is consistent with the results of \citet{mac91} and \citet{cox12}.
Since the shapes of the model curves are insensitive to $i$, we are unable to match the observed positions of $B$ and $C$.  If we force  the model curves to pass through $B$ and $C$ (see models 3 and 4), then the model boundary curves deviate from the position of $A$.  Nevertheless, our models suggest that the $v_*/\sqrt{n_{\rm ISM}}$ value is probably in the range of 15--50\,km\,s$^{-1}$\,cm$^{1.5}$.
For example, in model 1, the number density of H in the ISM would be 0.3\,cm$^{-3}$ if the star is moving at a speed of 10 km s$^{-1}$.

Below we will discuss the possible causes for the inability of Wilkin's expression in fitting the observations.
If the post-shock material is adiabatic, a termination shock and a bow shock are expected to form
inside and outside the halo, respectively. In this case, the boundary of the halo is the discontinuity
between stellar and interstellar material (i.e. astropause). These structures have been 
detected in the shell of IRC+10216 \citep{sah10}.  One may have an impression from Fig.~\ref{f1} that
the  NE bow-shaped filament detected by {\it GALEX} and the NE arc are located outside
and inside the halo respectively, and perhaps they represent the bow shock and the termination shock.
This scenario, however, is unlikely  because the surface brightnesses of the three components
are significantly different from those observed in IRC+10216 and predictions of
simulations. Moreover, the NE bow-shaped filament and arc have very faint counterparts on opposite
sides of the central star (see the right panel of Fig.~\ref{f1}), suggesting that they
were likely developed through different ejection events.  The appearance of the NE arc is
similar to that of the brightening shell as predicted by the simulations of \citet{war07},
probably suggesting that it was formed by the interaction between a fast wind and the early AGB wind.  
Furthermore, the appearances of the NE arc and its SW counterpart suggest that
it is likely to be a debris of the ring, a feature that has been frequently detected in halos of 
PNs \citep{cor04}.

Equations 1 and 2 are steady-state solutions. In the simulations of Betelgeuse's bow shock,
\citet{moh12} found that it takes a long time ($\ge 30\,000$\,yr) for the bow shock 
to reach a steady state, before which the $R(90^\circ)$/$R(0^\circ)$ ratio increases with time
from 1 to $\sqrt{3}$. Assuming that the symmetric axis of the halo of NGC\,7293 lies in the sky plane
(i.e. $i=0^\circ$), we obtain $R(90^\circ)$/$R(0^\circ)=1.32$. 
Therefore, a possible explanation for the mismatch in the model and observation results is
that the halo has not reached a steady state even though it is very old.

In addition, Wilkin's solutions are based on the assumption of a constant stellar wind, which is applicable for the early AGB stage. However, as the star evolves up the AGB, both the mass-loss rate and wind velocity increase in magnitude.  A jump in the stellar wind velocity in the post-AGB phase of evolution will cause interactions among stellar winds, which will also have an effect on the shape of the halo boundary.
The shape of the halo can also be affected by additional mechanisms such as magnetic fields, which 
may distort the halo boundary.  \citet{van11} performed simulations including  drag force between dust and gas. They found that depending on sizes of dust grains, infrared observations of bow shocks may not represent the morphologies of gas. The most likely explanation of the discrepancy between the model and the observed halo morphology is a latitude dependence of the density distribution of the AGB wind.  AGB winds are known to develop non-spherically symmetric outflows near the end of AGB evolution, and this can easily change the shape of the interaction region.  This equatorial enhanced mass loss have been attributed to binary central stars, star-planet interaction, star spots, and other mechanisms, but the exact cause is yet to be identified.  Another possible complication is the angular dependence of the later-developed fast wind (bipolar outflow), which can distort the density distribution of the AGB envelope.  Further numerical simulations including multiple stellar winds will be needed.  Additional variables such as time-dependence and angle-dependence of outflows need to be investigated.  The observed halo morphology presented here will serve as useful constraints to these models.

Recently, \citet{ram12} discovered an outer halo around the young PN IC\,418 based on the {\it WISE} images.
Unlike that of NGC\,7293, the halo of IC\,418 does not exhibit a cometary shape and has a well defined boundary, suggesting 
that it has not been striped. Its elliptical appearance indicates that the density and velocity AGB winds are indeed 
angle-dependent. \citet{vil03} performed 2-D numerical simulations of the PN-ISM interaction considering variable AGB mass-loss rates.
The shape of the NGC\,7293 halo is qualitatively similar to their results. The subsequent 3-D simulations by \citet{war07}
predicted four stages of the PN-ISM interaction. According to their classification scheme, this evolved PN is in the first
stage when the main nebula is unaffected by the interaction. \citet{war07b} showed that the Kelvin-Helmholtz instabilities at the shock front can lead to vortices in the wakes of AGB stars. Although we do not see unambiguous vortex-like structures in the halo of NGC\,7293,
the ``SW arc'' seemingly resembles the vortices reproduced by the models of \citet{war07b}. Through theoretical analysis, \citet{dga98}
found that because of the Rayleigh-Taylor instability the PN halo can be fragmented by the flow of ISM material. This phenomenon has been discovered in the PN NGC\,6772 \citep{ram09}. We note that the halo appearance of NGC\,7293 highly resembles to that of NGC\,6772. Both show bright rimmed region in the shock head, fan-like material distribution in the tail, and ray-like enhancements in the interior. It follows that the Rayleigh-Taylor disruption is also present in the halo of NGC\,7293. NGC\,7293 is presumably more evolved than those PNs with halos investigated by \citet{ram09}. We do not find tight relations between the properties of the halos and the ages of their host PNs. This should be ascribed to different environments surrounding these PNs which play an important role in the PN-ISM interaction processes.

\section{CONCLUSIONS}

From the imaging data of the {\it WISE} all-sky survey we discover a mid-infrared halo with a diameter of $\sim40'$ around NGC\,7293. The halo is totally invisible at optical wavelengths and emits light with a peak intensity near 12\,$\mu$m. The morphology of the halo suggests that it has been shaped by wind-ISM interaction, and the NE arc is brightened through wind-wind interaction. We find that the shape of the halo boundary cannot be explained by the analytic solutions of \citet{wil96} under the thin shock-shell and steady-state approximations.  This suggests that mass loss during the progenitor AGB phase has a more complicated history due to time- and/or angular-dependent outflows.

PN systems are the results of dynamical interactions between the halo (remnants of AGB wind) with the interstellar medium, as well as the interactions between fast outflows with the remnants of the AGB wind (interacting winds).  The quantitative information derived from this study is useful for the understanding of both types of dynamical interactions.  NGC\,7293, being the PN closest to us, is a valuable laboratory for the study of dynamic processes at work in the astrophysical environment.

\acknowledgments

We thank Sze Ning Chong for helpful discussions. We also thank the anonymous referee for helpful comments. This paper makes use of data products from the {\it Wide-field Infrared Survey Explorer}, which is a joint project of the University of California, Los Angeles, and the Jet Propulsion Laboratory/California Institute of Technology, funded by the National Aeronautics and Space Administration. Support for this work was provided by the Research Grants Council of the Hong Kong under grants HKU7031/10P and the Seed Funding Programme for Basic Research in HKU (200909159007).

\begin{deluxetable}{cccc}
\tablecaption{The Modeling Results$^a$. \label{t1} }
\tablewidth{0pt}
\tablehead{
\colhead{Model}&
\colhead{$i$}&
\colhead{$R_0$} &
\colhead{$v_*/\sqrt{n_{\rm ISM}}$} \\
&
\colhead{($\,^\circ$)}&
\colhead{($10^{18}$cm)}&
\colhead{(km\,s$^{-1}$\,cm$^{1.5}$)}\\
}
\startdata
1&  0  & 2.78 & 17.2 \\
2&  60 & 1.80 & 26.5 \\
3&  0  & 2.13 & 22.3 \\
4&  60 & 1.44 & 33.0 \\
\enddata
\tablenotetext{a}{Assuming $\dot M=10^{-5}$\,$M_\sun$\,yr$^{-1}$, $v_{\rm w}=10$\,km\,s$^{-1}$,
and the dimensionless mean molecular weight of 1.33.}
\end{deluxetable}

\end{CJK*}

\end{document}